\documentclass[aps,twocolumn,pra,superscriptaddress,amsmath,amssymb]{revtex4-1}
\usepackage{graphicx,color}
\usepackage{epstopdf}
\usepackage{epsfig}
\usepackage{mathrsfs}
\usepackage[breaklinks=true]{hyperref}

\newcommand{\beq}{\begin{equation}}
\newcommand{\eeq}{\end{equation}}
\newcommand{\bqa}{\begin{eqnarray}}
\newcommand{\eqa}{\end{eqnarray}}

\definecolor{green}{rgb}{0.00,0.50,0.00}

\begin{document}

\title{The Global versus Local Hamiltonian Description \\
of Quantum Input-Output Theory}
\author{John Gough} \email{jug@aber.ac.uk}
\date{\today}

\begin{abstract}
The aim of this paper is to derive the global Hamiltonian form for a quantum system and bath, or more generally a quantum network with multiple quantum input field connections, based on the local descriptions. We give a new
simple argument which shows that the global Hamiltonian for a single Markov component arises as the singular perturbation of the free translation operator. We show that the Fermi analogue takes an equivalent form provided
the parity of the coefficients is correctly specified. This allows us to immediately extend the theory of quantum feedback networks to Fermi systems.
\end{abstract}

\affiliation{Aberystwyth University, Aberystwyth, SY23 3BZ, Wales, United Kingdom}

\maketitle
\section{Introduction}

The quantum stochastic calculus was introduced by Hudson and Parthasarathy \cite{HP} in 1984 as a framework to construct explicit dilations of quantum dynamical evolutions (semigroups of completely positive norm-continuous
identity preserving maps) generalizing the usual It\={o} theory. Here the system Hilbert space is tensored with a Fock space over $L_{\mathbb{C}^{n}}^{2}(\mathbb{R})$ where $n$ enumerates the number of input noise channels. In particular, they showed that the most general form of a quantum unitary stationary evolution was obtained as the solution to an It\={o} quantum stochastic differential equation of the form 
\begin{eqnarray}
dU(t) &=&  \bigg\{ (S_{ij}-\delta _{ij})d\Lambda _{ij}(t)+L_{i}dB_{j}(t)^\ast 
 \nonumber \\
  &&-L_{i}^{\ast }S_{ij}dB_{j}(t) -(\frac{1}{2}L_{i}^{\ast }L_{i}+iH)dt  \bigg\} \, U(t).
\label{Ito qsde}
\end{eqnarray}
(implied sum of repeated indices over the range $1, \cdots, n$) with $S=[S_{jk}]$ unitary on $\mathfrak{h} \otimes \mathbb{C}^n$, $L=[L_j]$ arbitrary, and $H$ self-adjoint. These type of models are commonly referred to as following the $SLH$ formalism, see Section \ref{sec:QSDE} for details.

It was quickly realized \cite{Acc90} that the quantum evolutions could in fact be interpreted as a singular perturbation of the free dynamics corresponding to the second quantization of the shift along $L_{\mathbb{C}^{n}}^{2}(\mathbb{R})$. In particular, one could take the line $\mathbb{R}$ to physically be an infinite transmission line along which Bose quanta are propagating at constant speed, and interacting with the system (located at the origin) instantaneously as they pass through. It was a long standing program to find the form of the associated Hamiltonian, which when viewed in the interaction picture with respect to the free shift dynamics, gave the general quantum open dynamics, as well as a large class of classical stochastic models. This was first done by Chebotarev \cite{Cheb97} for the case of commuting coupling coefficients, and by Gregoratti \cite{Gregoratti}
for the general bounded operator case. The requirement of boundedness was later dropped \cite{QG}. The domain of the Hamiltonian is determined through a boundary condition at the origin on the vectors. 

The form of the Hamiltonian,  $K$, obtained in \cite{Cheb97} and \cite{Gregoratti} is given by 
\begin{eqnarray}
-i K\Psi =-i\tilde{K}_{0}\Psi -(\frac{1}{2}L_{i}^{\ast }L_{i}+iH)\Psi
-L_{i}^{\ast }S_{ij}b_{j}( 0^{+}) \Psi , \nonumber \\
\label{eq:Greg_Ham}
\end{eqnarray}
where 
\begin{eqnarray}
\tilde{K}_{0} &=& \left( \int_{-\infty }^{0}+\int_{0}^{+\infty } \right)
b_{j}( x) ^{\ast } \left( i\frac{\partial }{\partial x}\right) b_{j}( x)  \, dx, 
\label{eq:K_tilde_0}
\end{eqnarray}
is the Hamiltonian corresponding to free propagation of the external noise field (basically its the generator of translation, that is, the second quantisation of the momentum operator with the origin removed), and $\Psi$ is taken to belong to a domain of suitable functions satisfying the boundary condition 
\begin{eqnarray}
b_{i}( 0^{-}) \Psi =L_{i}\Psi +S_{ij}\,b_{j}( 0^{+}) \Psi .
\label{eq:BC}
\end{eqnarray}
with the suitable functions in question being those on the joint system and Fock space that are in the domain of the free translation along the positive and negative axis (excluding the vertex at the origin), and in the domain of the one-sided annihilators $b_i (0^\pm )$. 

The expression for the associated Hamiltonian looks strange because it is asymmetric in the formal creation and annihilation operator densities $b_j$ and $b_j^\ast$, and this makes the interpretation non-obvious for physicists. One of the aims of this paper is to show that there is a symmetric form of the Hamiltonian which is explicitly symmetric, and which is equivalent to the form (\ref{eq:Greg_Ham}) once the boundary condition (\ref{eq:BC}) is
taken into account.

We remark that a similar approach has been pioneered by von Waldenfels \cite{vonWaldenfels} which also exploits a quantum white noise formulation, the corresponding kernel calculus (that is, a Bose version of the Berezin calculus due to Maassen), and an explicit construction of deficiency spaces needed to then construct the self-adjoint Hamiltonian. In particular he succeeds in calculating the resolvent of the Hamiltonian. The approach outlined in this note is however more concise.

We shall refer to this situation of a single system interacting in a singular (=Markov) fashion with a quantum field moving along a transmission line as a \textit{local} Hamiltonian model. More generally we can consider several systems at various points on the transmission line, or more generally a network of transmission lines with systems at the vertices. The theory of quantum feedback networks has been developed recently to study such quantum mechanical systems connected by various arrangements of quantum field inputs \cite{GoughJamesCMP09}. Here the notion of a \textit{global} Hamiltonian was introduced in \cite{GoughJamesCMP09} in order to construct tractable models of a quantum feedback network. For a single component, this reduces to the Hamiltonian obtained by Chebotarev \cite{Cheb97} and Gregoratti \cite{Gregoratti}, and we present a simple derivation of this object in section \ref{sec:SingPert} below. The network theory has been applied to quantum optics with a view to developing closed-loop quantum control systems \cite{GoughJamesIEEE09}- \cite{N10}.

The separate components are modeled as quantum open systems \cite{HP}, see also \cite{partha}, or equivalently as quantum input-output systems \cite{Gardiner}. In a network, connections are made by feeding the output of one
system in as input to another, or even to the same component again. In a physically realistic model, this will involve time-lags, however it often useful to take the limit of instantaneous connections. Moreover the components themselves are assumed to be capable of scattering inputs, in particular acting as beam-splitters.

The has been interest amongst theoreticians in recent years in using solid-state devices as in place of optical fields \cite{MilburnSunUpcroft00}-\cite{AF07}. In suitable regimes the models resemble the optical case with the obvious exception that the fields are now Fermionic rather than Bosonic.
In section \ref{sec:Fermi}, we give the Fermi analogue and show that the global Hamiltonian method applies equally well to this situation. Indeed, provided the various coupling terms meet the physical conditions regarding
Fermionic parity, the essential rules governing the construction of networks (the series product \cite{GoughJamesIEEE09}, the concatenation rule and feedback reduction rule \cite{GoughJamesCMP09}) are identical to the Bose
input case.

The outline of this paper is that we first review the background theory of open quantum systems adopting a quantum white noise convention. We make some tentative links with the theory of singular perturbations of unbounded below
Hamiltonians (which is of relevance to the free Hamiltonian generating the shift). In section \ref{sec:SingPert} we give an alternative rationale behind the form of the associated local Hamiltonian, and outline how this may be generalized to the case of a global Hamiltonian for a quantum feedback network. In section \ref{sec:Fermi}, we treat the extension to Fermi noise models, and indicate that the formal expressions should be the same as the Bose case provided that the coefficients have specific parity with respect to the $\mathbb{Z}_{2}$-grading of Fermi Fock space. Finally in section \ref{sec:network} we indicate how, in the limit of zero time delay in a network edge, we obtain a modular reduction in the models.

\subsection{Dynamical Perturbations}
Let $K$ be a Hamiltonian of the form
\begin{eqnarray*}
K=K_0 +\Upsilon,
\end{eqnarray*}
and define $V_0 (t)$ to be the unitary dynamics generated by the free Hamiltonian $K_0$, and $V(t)$ the perturbed unitary generated by $K$. We may move to the Dirac picture where we view $V$ as a perturbed dynamics with respect to the free dynamics of $V_{0}$, and to this end introduce the unitary transforming from to the interaction picture (sometimes referred to as the wave operator)
\begin{eqnarray}
U( t) =V_{0}( t) ^{\ast }\, V( t) .
\label{eq:wave_operator}
\end{eqnarray}
The pair $V_{0}( t) $ and $V( t) $ are strongly continuous one-parameter groups, that is $V_{0}(t+s)=V_{0}(t)V_{0}(s)$ and $V(t+s)=V(t)V(s)$, however, we family of wave operators satisfy the so-called \emph{cocycle property}
with respect to $V_0$ 
\begin{eqnarray}
U( t+s) =\Theta _{t}( U(s)) \, U(t),
\label{eq:cocycle}
\end{eqnarray}
where we encounter the free dynamics 
$$\Theta _{t}(x)=V_{0}(t)^{\ast }X\, V_{0}(t).$$
While the unitary groups satisfy the Schr\"{o}dinger equations
\begin{eqnarray*}
i\dot{V}_{0}(t)=K_{0}\, V_{0}(t),\quad i\dot{V}(t)=K\, V(t).
\end{eqnarray*}
we find that the wave operator satisfies the Dirac interaction picture equation
\begin{eqnarray*}
i\dot{U}( t) =\Upsilon (t) \, U(t)
\label{eq:Dirac}
\end{eqnarray*}
where the time-dependent Hamiltonian is 
\begin{eqnarray*}
\Upsilon (t)=\Theta _{t}(\Upsilon ).
\end{eqnarray*}

\bigskip

Conversely, suppose we are given a strongly continuous unitary group $V_0(t)$ with Hamiltonian generator $K_0$, as well as a strongly continuous cocycle $U(t)$ with respect to $V_0$. We may \textit{define} a unitary family $V(\cdot )$ by
\begin{eqnarray*}
V(t) = V_0 (t) \, U(t) ,
\end{eqnarray*}
which is just (\ref{eq:wave_operator}) rearranged. The fact that $U$ is a $V_0$-cocycle now implies that $V$ defined in this manner will be a unitary group -this follows in a purely algebraic way from (\ref{eq:cocycle}. Moreover, $V$ will be strongly continuous by construction.

According to Stone's Theorem, $V$ should then possess a Hamiltonian generator $K$. We say that $K$ is a regular perturbation of $K_{0}$ if the difference $\Upsilon =K-K_{0}$ defines an operator with dense domain on the Hilbert space. In this case, $U(t)$ will be strongly differentiable and we are lead to the interaction picture equation (\ref{eq:Dirac}).

In situations where $\Upsilon $ is not densely defined, however, we will have a singular perturbation. In this case, $U(t)$ will always be strongly continuous, but not generally strongly differentiable.

\subsection{Stochastic Evolutions driven by Classical Noise}
To understand better, let us consider some examples of a quantum system with a fixed Hilbert space $\mathfrak{h}$ of states, driven by classical noise. For instance, let $U(t)$ be given by 
$$ U(t) = e^{-iE \, W(t)}$$
where $W(t)$ is a Wiener process and $E$ is a self-adjoint operator.
We have the It\={o} rule that $(dW(t))^2 =dt$, so that the stochastic differential equation satisfied by
$V(t)$ is
\begin{eqnarray}
dU(t) = \left\{ -iE dW(t) - \frac{1}{2} E^2 dt \right\} \, U(t)
\end{eqnarray}
It is clear from the fact that $U(t)$ solves a stochastic differential equation and as such will not be strongly differentiable.

To see how this fits in to our discussions on perturbations, let us recall that the Wiener process is correctly defined on the sample space $\Omega_W = \mathscr{C}_0( \mathbb{R}_+ )$ of continuous paths parameterised by a continuous variable $t \geq 0$ starting at the origin at time $t=0$. The set of measurable subsets of paths are the cylinder sets, and a probability measure on these subsets of paths is given by Wiener measure $\mathbb{P}_W$.
Let $\omega$ be a Wiener path, then $W(t) \, (w) $ is the evaluation $\omega (t)$ - the coordinate of the path at time $t$! We then set  $ U(t, \omega ) = e^{-iE \, \omega (t)}$ as the evaluation of the random variable $U(t)$ when the outcome is a particular path $\omega \in \Omega_W$.

We may now view the model as being on the total Hilbert space $\mathfrak{h} \otimes \mathscr{H}$ where the system is now coupled to an environment whose Hilbert space is
\begin{eqnarray*}
\mathscr{H} = L^2 (\Omega_W , \mathbb{P}_W ).
\end{eqnarray*}
As is well-known $\mathscr{H}$ is isomorphic to Bose Fock space over $L^2 ( \mathbb{R}_+ ,dt)$: in other words, the Hilbert space on which we define a single quantum input process. In effect, we may think of $W(t)$ as the quadrature of a quantum field in the quantum input theory. We now introduce the time-shift $\Theta_t $ on the classical Wiener probability space as
\begin{eqnarray*}
\Theta_t ( W(s) ) \triangleq W(t+s),
\end{eqnarray*}
for all $t,s \geq 0$. We then have
\begin{eqnarray*}
U (t+s ) &=&  e^{-i E \, W(t+s) } \\
&=& e^{-iE \, [W(t+s) -W(s )] } \, e^{-i E \, W(t) }\\
&=& \Theta_s \left( e^{-iE \, [W(t) -W(0 )] } \right) \, e^{-i E \, W(t) } \\
&=& \Theta_t \left( U(s) \right) \,  U(t) ,
\end{eqnarray*}
which states that the family of unitaries $U(t)$ is a cocycle with respect to the free translation of the Wiener noise. A physical picture to have here is that of a quantum input field propagating down a semi-infinite wire and interacting with the system located at the origin of the wire, the interaction involves only a quadrature of the field - $W(t)$ - and the unitary will be diagonal with respect to this quadrature. Suppose the system is initiated in state determined by a density matrix $\varrho_0$, then averaging over the Wiener paths leads to the state
\begin{eqnarray*}
\varrho_t = \int_{\Omega_W} U(t, \omega ) \varrho_0 U (t, \omega )^\ast \, \mathbb{P} [d \omega ].
\end{eqnarray*}
From the It\={o} calculus we find the master equation
\begin{eqnarray*}
\dot{\varrho_t} = E \varrho_t E -\dfrac{1}{2} \{ E , \varrho_t \} .
\end{eqnarray*}

\bigskip

A similar theory can be developed for quantum jumps. Let $N(t)$ we the Poisson process with rate $\nu$ so that we have the rule $(dN(t))^p =dN(t)$ for $p=1,2,3,\cdots$ and $\mathbb{E} [dN(t)] = \nu \, dt$. Then a unitary process is defined by
\begin{eqnarray*}
U(t) = S^{N (t)} ,
\end{eqnarray*}
where $S$ is a fixed unitary on the Hilbert space $\mathfrak{h}$ of the system. We obtain the stochastic differential equation
\begin{eqnarray*}
dU(t) = (S-I) U(t) \, dN(t), \quad U(0)=I,
\end{eqnarray*}
and due to the unitary kicks by the operator $S$ occurring at random times, the process is again strongly continuous, but not strongly differentiable. The time shift map on Poisson sample paths may once again be seen as a free dynamics, with $U(t)$ once more being a cocycle with respect to this free dynamics. The corresponding master equation may be obtained by averaging over all Poisson paths and this corresponds to
\begin{eqnarray*}
\dot{\varrho_t} = \nu \{S \varrho_t S^\ast  -\varrho_t \} .
\end{eqnarray*}

\subsection{Classical SDEs}
To see why this result is astonishing, let us consider a purely classical problem of a particle with position $\mathbf{x}$ moving in a deterministic vector field $\mathbf{v} ( \mathbf{r})$, but subject to classical Wiener noise.

We consider the classical diffusion process defined as the solution to the
stochastic differential equation
\begin{equation}
d\mathbf{x}\left( t\right) =\mathbf{v}\left( \mathbf{x}\left( t\right)
\right) \,dt+\mathbf{\sigma }\left( \mathbf{x}\left( t\right) \right)
\,dW\left( t\right) .  \label{eq:SDE}
\end{equation}

The dynamics can be obtained from a quantum mechanical model by taking $%
\mathbf{q}$ to be position operator in the Schr\"{o}dinger picture and
introducing a conjugate momentum $\mathbf{p}$.  Consider the stochastic unitary process determined by the SDE

\begin{eqnarray}
dU(t) = \left\{ -iE \, dW(t) - \frac{1}{2} E^2 \, dt -iH \, dt \right\} \, U(t)
\end{eqnarray}
with the Hamiltonian 
\begin{equation*}
H=\frac{1}{2}\mathbf{p}.\mathbf{w}\left( \mathbf{q}\right) +\frac{1}{2}%
\mathbf{w}\left( \mathbf{q}\right) .\mathbf{p}
\end{equation*}
and where we set
\begin{equation*}
E=\frac{1}{2}\mathbf{p}.\mathbf{\sigma }\left( \mathbf{q}\right) +\frac{1}{2}%
\mathbf{\sigma }\left( \mathbf{q}\right) .\mathbf{p}
\end{equation*}
In the Heisenberg picture, $\mathbf{q}_{t}=U\left( t\right) ^{\ast }\mathbf{q%
} \, U\left( t\right) $ is then seen to satisfy the same SDE (\ref{eq:SDE}) as $%
\mathbf{x}\left( t\right) $ with drift vector 
\begin{equation*}
v_{i}=w_{i}+\frac{1}{2}\frac{\partial \sigma _{i}}{\partial x^{j}}\sigma
_{j}.
\end{equation*}
The vector $\mathbf{w}$ is known as the Stratonovich drift, while $\mathbf{v}
$ is the It\={o} drift.

The somewhat surprising conclusion is that every classical diffusion process (or jump process,
and by extension any process driven by independent increment processes) can be obtained as the 
interaction picture evolution of a commuting set of position observables arising as 
a singular perturbation.

\subsection{Quantum Stochastic Evolutions}

As we have mentioned, Quantum stochastic evolutions were introduced in \cite{HP}, and physical models describing quantum optics systems driven by quantum input processes $b_{i}(t)$ were independently given by
 \cite{Gardiner}. the latter theory is more formal, but it is useful to view these quantum input processes as singular operator densities. 

\subsubsection{Fock Space}
Let us fix the appropriate Hilbert space to be the Fock space $\mathfrak{F}$ over the one-particle space $\mathbb{C}^{n}\otimes L^{2}(\mathbb{R})$. For $\Psi \in \mathfrak{F}$, we have a well-defined amplitude $\langle \tau _{1},i_{1};\cdots ;\tau _{m},i_{m}|\Psi \rangle $ which is completely symmetric under interchange of the $m$ pairs of labels $(\tau _{1},i_{1}),\cdots ,(\tau _{m},i_{m})$, and this represent the amplitude to have $m$ quanta with a particle of type $i_{1}$ at $\tau _{1}$, particle of type $i_{2}$ at $\tau _{2}$, etc. We have the following resolution of identity on $\mathfrak{F}$:
\begin{eqnarray*}
I &=& \sum_{m=0}^{\infty }(\int d\tau _{1}\cdots d\tau
_{m})(\sum_{i_{1}=1}^{n}\cdots \sum_{i_{m}=1}^{n}) \\
&& \times |\tau
_{1},i_{1};\cdots ;\tau _{m},i_{m}\rangle \langle \tau _{1},i_{1};\cdots
;\tau _{m},i_{m}|.
\end{eqnarray*}
The annihilator input process $b_{i}(t)$ is then realized as 
\begin{eqnarray*}
\langle \tau _{1},i_{1};\cdots ;\tau _{m},i_{m}|b_{i}(t)\Psi \rangle \\
=\sqrt{m+1}\,\langle t,i;\tau _{1},i_{1};\cdots ;\tau _{m},i_{m}|\Psi \rangle .
\end{eqnarray*}

The annihilation operators, together with their formal adjoints the creator operators $b_{i}(t)^{\ast }$, satisfy the singular canonical commutation relations 
\begin{eqnarray}
\lbrack b_{i}( t) ,b_{j}(s)]=\delta _{ij}\delta ( t-s) .  \label{sing CCR}
\end{eqnarray}

\subsubsection{The Time Shift for Fields on Fock Space}

We may define an operator $K_0$ on the Fock space by 
\begin{eqnarray}
K_{0}=\sum_{j=1}^{n}\int_{-\infty }^{\infty }dt\,b_{j}^{\ast }(t)  \left( i \frac{%
\partial }{\partial t} \right) b_{j} (t) \label{K_0}
\end{eqnarray}
and this is the second quantization of the one-particle momentum operator $i\frac{\partial }{\partial t}$. (Note that $t$ is not physical time, but distance along the spatial axis measured in arc time: that is if the field
quanta are modelled as propagating down the $x$-axis with speed $c$, then $t \equiv x/c$ is the time to reach the system located at the origin. Causality comes into play as $b_j(t)$ describes the field that arrives at the system at times $t$.) The Hamiltonian $K_0$ is self-adjoint operator, and it generates the unitary group $V_{0}(t)=e^{-itK_{0}}$  giving the time shift (propagation of quanta down the axis): 
\begin{eqnarray*}
\langle \tau _{1},i_{1};\cdots ;\tau _{m},i_{m}|&V_{0}(t)&\Psi \rangle \\
&=&\langle \tau _{1}+t,i_{1};\cdots ;\tau _{m}+t,i_{m}|\Psi \rangle .
\end{eqnarray*}
The free dynamics $\Theta _{t}(\cdot )=V_{0}(t)^{\ast }(\cdot )V_{0}(t)$ will then translate the input processes: 
\begin{eqnarray*}
\Theta _{\tau }(b_{i}(t))=b_{i}(t+\tau ),\;\Theta _{\tau }(b_{i}^{\ast
}(t))=b_{i}^{\ast }(t+\tau ).
\end{eqnarray*}

\subsubsection{The Local Hamiltonian}

Let us fix a system space $\mathfrak{h}$ and consider a singular perturbation on $\mathfrak{h}\otimes \mathfrak{F}$\ of the form 
\begin{eqnarray}
\Upsilon =E_{ij}b_{i}^{\ast }(0)b_{j}(0)+E_{i0}b_{i}^{\ast
}(0)+E_{0j}b_{j}(0)+E_{00},  \label{Upsilon}
\end{eqnarray}
with $E_{ij}^{\ast }=E_{ji}$, $E_{i0}^{\ast }=E_{0i}$ and $E_{00}^{\ast }=E_{00}$. We obtain a time-dependent Hamiltonian by means of the time shift 
\begin{eqnarray*}
\Upsilon (t) &=& \Theta _{t}(\Upsilon ) \\
&=& E_{ij}b_{i}^{\ast
}(t)b_{j}(t)+E_{i0}b_{i}^{\ast }(t)+E_{0j}b_{j}(t)+E_{00}.
\end{eqnarray*}
The solution to the formal equation $\dot{U}(t)=-i\Upsilon (t)U(t)$, with initial condition $U( 0) =I$, may be expresses as the Dyson series expansion 
\begin{eqnarray*}
U( t) =\sum_{n=0}^{\infty }(\frac{1}{i})^{n}\int_{\Delta _{n}( t) }\Upsilon
( \tau _{n}) \cdots \Upsilon ( \tau _{1})
\end{eqnarray*}
where we encounter integration over the simplices $\Delta _{n}( t) $ of times $t\geq \tau _{n}>\cdots >\tau _{1}\geq 0$. The formal series may be rewritten as the chronologically ordered exponential which we may denote as
\begin{eqnarray}
U( t) =\vec{T}\exp \frac{1}{i}\int_{0}^{t}\Upsilon (\tau )d\tau .
\label{U chronological}
\end{eqnarray}
Note that the notation is identical to that used in quantum field theory, however, the Hamiltonian $\Upsilon (t)$ is more singular in this case. Nevertheless we find that many algebraic identities carry over immediately, and in particular we observe that $U(t)$ is a $\Theta $-cocycle since  
\begin{eqnarray*}
U(t+s) &=&\vec{T}\exp \frac{1}{i}\int_{t}^{t+s}\Upsilon (\tau )d\tau \;\vec{T%
}\exp \frac{1}{i}\int_{0}^{t}\Upsilon (\tau )d\tau \\
&=&\Theta _{t}( \vec{T}\exp \frac{1}{i}\int_{0}^{s}\Upsilon (\tau )d\tau ) \;%
\vec{T}\exp \frac{1}{i}\int_{0}^{t}\Upsilon (\tau )d\tau \\
&=&\Theta _{t}( U(s) )\,U(t).
\end{eqnarray*}

\subsubsection{Wick Ordered Form}

Let us briefly indicate how to convert $U(t)$ to Wick order \cite{Gough97}, \cite{Gough99} \cite{GoughWong-Zakai}. Starting from the integro-differential equation $U(t)=1-i\int_{0}^{t}\Upsilon (s)U(s)ds$, we have 
\begin{eqnarray*}
[ b_{i}(t),U(t)] &=&-i\int_{0}^{t}[b_{i}(t),\Upsilon (s)]U(s)ds \\
&=&-i\int_{0}^{t}\delta _{ij}\delta (t-s)\{E_{jk}b_{k}(t)+E_{j0}\}U(s) \\
&=&-\frac{i}{2}E_{ij}b_{j}(t)U(t)-\frac{i}{2}E_{i0}U(t).
\end{eqnarray*}
Here we assumed that $[b_{i}(t),U(s)]=0$ for $t>s$, since $U(s)$ depends only on the noise up to time $s$. We also adopted the convention that the $\delta $-function contributes on half-weight due to the upper limit of the
integral. This implies that $b_{i}(t)U(t)=[(1+\frac{i}{2}E)^{-1}]_{ij}[U(t)b_{j}(t)-\frac{i}{2}E_{j0}U(t)]$ and we may use this to set the equation $\dot{U}(t)=-i\Upsilon (t)U(t)$ to Wick order, to obtain 
\begin{eqnarray}
\dot{U}(t) &=& b_{i}(t)^{\ast }(S_{ij}-\delta_{ij})U(t)b_{j}(t)+b_{i}(t)^{\ast
}L_{i}U(t) \nonumber \\
&& -L_{i}^{\ast }S_{ij}U(t)b_{j}(t)-(\frac{1}{2}L_{i}^{\ast
}L_{i}-iH)U(t),  \label{Wick QSDE}
\end{eqnarray}
where $S=[S_{ij}]$ is the Cayley transform $E=[E_{ij}]$, 
\begin{eqnarray}
S=\frac{1-\frac{i}{2}E}{1+\frac{i}{2}E}  \label{S}
\end{eqnarray}
and therefore unitary, while 
\begin{eqnarray}
L_{i} &=& i\left[ \frac{1}{1+\frac{i}{2}E}\right] _{ij}E_{j0},\nonumber \\
H &=& E_{00}+\frac{1%
}{2}E_{0i}\left[ \mathrm{Im} \frac{1}{1+\frac{i}{2}E}\right] _{ij}E_{j0}
\label{L's}
\end{eqnarray}
with $H$ self-adjoint.

\subsubsection{It\={o} Quantum Stochastic Differential Form}
\label{sec:QSDE}
Introducing integrated fields 
\begin{eqnarray}
B_{i}(t)^{\ast } &=&\int_{0}^{t}b_{i}(\tau )^{\ast }d\tau
,\;B_{i}(t)=\int_{0}^{t}b_{i}(\tau )d\tau , \\
\Lambda _{ij}(t) &=&\int_{0}^{t}b_{i}(\tau )^{\ast }b_{j}(\tau )d\tau ,
\label{Bose processes}
\end{eqnarray}
called the creation, annihilation and gauge processes, respectively, it is possible to define quantum stochastic It\={o} integrals with respect to these fields. Conditions for the existence and uniqueness of solutions is given and, for a fixed system space $\mathfrak{h}$ are given in \cite{HP}. The equation $(\ref{Wick QSDE})$ is readily interpreted as the It\={o} quantum stochastic differential equation 
\begin{eqnarray}
dU(t) =  \{ (S_{ij}-\delta _{ij})d\Lambda _{ij}(t)+L_{i}dB_{j}(t)^\ast
 \nonumber \\
 -L_{i}^{\ast }S_{ij}dB_{j}(t) -(\frac{1}{2}L_{i}^{\ast }L_{i}+iH)dt  \} U(t).
\label{Ito qsde}
\end{eqnarray}
With the already deduced conditions that $S=[S_{ij}]$ be a unitary matrix and $H$ self-adjoint, $(\ref{Ito qsde})$ gives the general equation satisfied by a unitary adapted quantum stochastic process. The rather non-Hamiltonian appearance of the equation is a result of the fact that we have the following non-trivial products of It\={o} differentials: 
\begin{eqnarray}
dB_{i}dB_{j}^{\ast } &=&\delta _{ij}dt,\quad dB_{i}d\Lambda _{jk}=\delta
_{ij}dB_{k}, \\
d\Lambda _{ij}dB_{k}^{\ast } &=&\delta _{jk}dB_{i}^{\ast },\quad d\Lambda
_{ij}d\Lambda _{kl}=\delta _{jk}d\Lambda _{il}.  \label{Ito table}
\end{eqnarray}
The It\={o} convention for differentials, that is where the increment $dY(t)$ is taken as the future pointing increment $Y(t+dt)-Y(t)$ and $X(t)dY(t)$ is understood at the infinitesimal level as $X(t)[Y(t+dt)-Y(t)]$. As an
alternative we could use the Stratonovich convention which is to take the midpoint rule 
\begin{eqnarray*}
X(t)\circ dY(t) &=& X(t+\frac{1}{2}dt)[Y(t+dt)-Y(t)]\\
&\equiv & X(t)dY(t)+\frac{1}{2}%
dX(t)dY(t).
\end{eqnarray*}
The Stratonovich quantum stochastic differential equation corresponding to $(\ref{Ito qsde})$ is then

\begin{eqnarray*}
dU=-i\{ E_{ij}d\Lambda_{ij} +E_{i0}dB_{i}^{\ast }+E_{0j}dB_{j}+E_{00}dt\}
\circ U.
\end{eqnarray*}

\subsubsection{The Langevin Equations}

For a system operator $X$, we set 
\begin{eqnarray*}
j_{t}( X) =U( t) ^{\ast }XU( t) .
\end{eqnarray*}

By reference to the quantum It\={o} rules, we see that, for $S=I$, we have
that 
\begin{eqnarray*}
dj_{t}(X) &=&U(t)^{\ast }XdU(t)+dU(t)^{\ast } \, X \otimes I \, U(t)\\
&& +dU(t)^{\ast }XdU(t) \\
&=&j_{t}([X,L_{i}])dB_{i}(t)^{\ast }+j_{t}([L_{i}^{\ast },X])dB_{i}(t)\\
&+& j_{t}(\mathcal{L}X)dt
\end{eqnarray*}
where we encounter the Lindblad generator 
\begin{eqnarray*}
\mathcal{L}X=\frac{1}{2}L_{i}^{\ast }[X,L_{i}]+\frac{1}{2}[L_{i}^{\ast },X%
]L_{i}-i[X,H].
\end{eqnarray*}

For $S\neq I$, we have the general Langevin equation 
\begin{eqnarray*}
dj_{t}(X) &=&j_{t}(S_{ki}^{\ast }XS_{kj}-\delta _{ij}X)d\Lambda _{ij}(t) \\
&&+j_{t}(S_{ji}^{\ast }[X,L_{j}])dB_{i}(t)^{\ast }+j_{t}([L_{i}^{\ast
},X]S_{ij})dB_{j}(t)\\
&+& j_{t}(\mathcal{L}X)dt.
\end{eqnarray*}

\subsubsection{Input-Output Relations}

The output processes $B_{i}^{\mathrm{out}}(t)$ are defined by the identity 
\begin{eqnarray}
B_{i}^{\mathrm{out}}(t)=U(t)^{\ast }\, I \otimes  B_{i}(t)\, U(t),  \label{IO relation}
\end{eqnarray}
which in differential form takes the following form 
\begin{eqnarray}
dB_{i}^{\mathrm{out}}(t)= j_t (S_{ij} ) \, dB_{j}(t)+  j_t (L_{i} ) \, dt.  \label{IO diff}
\end{eqnarray}
We see that the differential of the output is a unitary rotation of the input by the matrix $S$ in the interaction picture, plus a drift term corresponding to the coupling $L$ in the interaction picture. 
From the defining relation (\ref{IO relation}) we see that the output
processes satisfy the canonical commutation relations. 

The local version of this relation is then 
\begin{eqnarray}
b_{i}^{\mathrm{out}}(t)= j_t (S_{ij} ) \, b_{j}(t)+ j_t ( L_{i}).
\label{IO local}
\end{eqnarray}
It is clear that the local form (\ref{IO local}) is structurally similar to the boundary conditions (\ref{eq:BC}). In this sense the boundary conditions do indeed capture the input-output relations as the fields
propagate across the boundary (the system!) at the origin.

\subsection{Unitary QSDEs as Singular Perturbations}

The quantum stochastic process $U(t)$ define by either formally by $( \ref{Wick QSDE}) $ or mathematically as a solution to $( \ref{Ito qsde}) $, will be strongly continuous, but due to the presence of the noise fields $dB_{i}^{\ast },dB_{j}$ and $d\Lambda _{ij}$ will not be strongly differentiable. Indeed $U(t)$ is a singular perturbation of the generator of the time-shift $( \ref{K_0}) $, and the perturbation is formally given by the local interaction $\Upsilon $ given by $( \ref{Upsilon}) $.

We remark that nevertheless $U(t)$ is a $\Theta $-cocycle, and that we can then define a strongly continuous unitary group, $V( t) $, by 
\begin{eqnarray*}
V\left( t\right) =\left\{ 
\begin{array}{cc}
V_{0}\left( t\right) U\left( t\right) , & t\geq 0; \\ 
U\left( -t\right) ^{\ast }V_{0}\left( t\right) , & t<0.
\end{array}
\right.
\end{eqnarray*}
Our goal is to describe the infinitesimal generator $K$ of $V(t)$.

This has been a long standing problem \cite{Acc90}, and we recall briefly the path that lead to the form presented in (\ref{eq:Greg_Ham}) with boundary conditions (\ref{eq:BC}). The major breakthrough came
in 1997 when A.N. Chebotarev solved this problem for the class of quantum stochastic evolutions satisfying Hudson-Parthasarathy differential equations with bounded \textit{commuting} system coefficients \cite{Cheb97}. His insight was based on scattering theory of a one-dimensional system with a Dirac delta potential, say, with formal Hamiltonian 
\begin{eqnarray*}
k=i\partial +E\delta
\end{eqnarray*}
describing a one-dimensional particle propagating along the negative $x$-axis with a delta potential of strength $E$ at the origin. (In Chebotarev's analysis the $\delta $-function is approximated by a sequence of regular
functions, and a strong resolvent limit is performed.) As the coefficients were assumed to commute, he was able to perform a simultaneous diagonalisation of these operators and treat the problem in a class of states parameterised by the eigenvalue coordinate.  The mathematical techniques used in this approach were subsequently generalized by Gregoratti 
\cite{Gregoratti} to relax the commutativity condition. More recently, the analysis has been further extended to treat unbounded coefficients \cite{QG}.

Independently, several authors have been engaged in the program of describing the Hamiltonian nature of quantum stochastic evolutions by interpreting the time-dependent function $\Upsilon \left( t\right) $ as
being an expression involving quantum white noises satisfying a singular CCR \cite{Gough97},\cite{Gough99}, \cite{AVL}, \cite{vonWaldenfels}. 

An interesting historical point is that Chebatorev noted that, since traditional scattering techniques were based on perturbations of laplacain operators, they would not be immediately applicable to the situation here where the generator of the free dynamics $k_{0}=i\partial $ is a first order differential operator, and not semi-bounded. However newer methods introduced by Albeverio and Kurasov \cite{AlKur97},\cite{AlKur99a},\cite{AlKurBook00} may be employed to construct self-adjoint extensions of such models in the situation considered here where we interpret the singular interaction as a $\delta $-perturbation viewed as a singular rank-one perturbation. 
We will show in the next section that this is the case here for a wave on a 1-D wire.

Our starting point will be the one-dimensional model considered by Chebotarev, though consider this as a problem of finding a suitable self-adjoint extension for the singular second quantized Hamiltonian and present a intuitive argument leading to the correct from of $K$.

\section{Global Hamiltonian as Singular Perturbation of the Time Shift Generator}

\label{sec:SingPert}

\subsection{Single quantum on a 1-D wire}

We begin it a model of a single quantum mechanical particle moving in one-dimension with free Hamiltonian 
\begin{eqnarray*}
H_{0}=-vp=iv\hbar \frac{\partial }{\partial x}.
\end{eqnarray*}

The evolution is just the translation of the wave-function at velocity $v$ along the negative $x$-axis: 
\begin{eqnarray*}
\langle x|\psi (t)\rangle =\psi (x+vt).
\end{eqnarray*}
For simplicity we take $\hbar =v=1$, so that $x$ is arc-time along the wire.

We shall consider the singular perturbation consisting of a $\delta $-kick at the origin: 
\begin{eqnarray*}
H=H_{0}+\epsilon \delta (x).
\end{eqnarray*}

For the particle coming in from the right, it will feel an impulse as it passes the origin which will have the nett effect of introducing a jump discontinuity. The singular part of the Schr\"{o}dinger equation, $i\dot{\psi }=H\psi $ is then 
\begin{eqnarray*}
i[\psi (0^{+})-\psi (0^{-})]+\epsilon \frac{\psi (0^{+})+\psi (0^{-})}{2},
\end{eqnarray*}
where we have the momentum impulse (proportional to the jump in $\Psi $ at the origin) and the average value picked out by the $\delta $-function. To obtain a self-adjoint extension, we argue that this term vanishes exactly,
and this implies the boundary condition  
\begin{eqnarray*}
\psi (0^{-})=s \psi (0^{+}).
\end{eqnarray*}
where 
\begin{eqnarray*}
s=\frac{1-{\frac{i}{2}}\epsilon }{1+{\frac{i}{2}}\epsilon }.
\end{eqnarray*}
At present, we are considering the class of square-integrable functions with a possible jump discontinuity at $x=0$ for which the derivative function exists away from zero and is again square-integrable. We note the following
integration by parts formula for functions $\phi ,\psi $ in this class: 
\begin{eqnarray*}
&& \int \phi ^{\ast }( i\frac{\partial }{\partial x}\psi ) =\\
&& \int ( i\frac{%
\partial }{\partial x}\phi ) ^{\ast }\psi -i\phi ^{\ast }( 0^{+}) \psi (
0^{+}) +i\phi ^{\ast }( 0^{-}) \psi ( 0^{-}) .
\end{eqnarray*}
Let us denote the bras $\langle 0^{\pm }| $ and $\langle \bar{0}| =\frac{1}{2}\langle 0^{+}| +\frac{1}{2}\langle 0^{-}| $ defined on this class by 
\begin{eqnarray}
\langle 0^{\pm }|\psi \rangle =\psi ( 0^{\pm }) ,\quad \langle \bar{0}|\psi
\rangle =\frac{\psi ( 0^{+}) +\psi ( 0^{-}) }{2}  \label{split delta}
\end{eqnarray}
then we have 
\begin{eqnarray*}
\langle \phi |i\frac{\partial }{\partial x}\psi \rangle =\langle i\frac{%
\partial }{\partial x}\phi |\psi \rangle +\langle \phi |j\psi \rangle
\end{eqnarray*}
where the jump term is 
\begin{eqnarray*}
j &=&-i| 0^{+}\rangle \langle 0^{+}| +i| 0^{-}\rangle \langle 0^{-}| \\
&\equiv &-i| 0^{+}-0^{-}\rangle \langle \bar{0}| +i| \bar{0}\rangle \langle
0^{+}-0^{-}| ,
\end{eqnarray*}
where $\langle 0^{+}-0^{-}| \psi \rangle =\psi ( 0^{+}) -\psi ( 0^{-}) $. The operator $i\frac{\partial }{\partial x}$, understood in the current distributional sense, is clearly not symmetric on this class of functions,
however we do have $\langle \phi |k_{0}\psi \rangle =\langle k_{0}\phi |\psi \rangle $ where 
\begin{eqnarray}
k_{0}=i\frac{\partial }{\partial x}+i| \bar{0}\rangle \langle 0^{+}-0^{-}| .
\label{little k}
\end{eqnarray}

From von Neumann's theory of self-adjoint extensions of symmetric operators, it is well known that all self-adjoint extensions of the momentum operator defined on the 1-dimensional line with the origin removed are determined by
a boundary condition $\psi (0^{-})=s\psi (0^{+})$ where $s$ is unimodular, as is the case here. Evidently, this captures in, a very simple setting, the type of scattering that we see in $( \ref{S}) $. We now show that this
problem can be second-quantized without too much difficulty. 

\subsection{Indefinite number of identical (Bose) quanta on a 1-D wire}

To deal with discontinuities at zero, we introduce the averaged noise 
\begin{eqnarray}
\bar{b}_{i}( 0) =\frac{1}{2}b_{i}( 0^{+}) +\frac{1}{2}b_{i}( 0^{-}) .
\end{eqnarray}

Following our remarks leading to (\ref{little k}), we split $ K_{0}=\int_{-\infty }^{+\infty }b( x) ^{\ast }i\frac{\partial }{\partial x}b( x) dx=\tilde{K}_{0}+J$ where
\begin{eqnarray}
\tilde{K}_{0} &=&\sum_{j=1}^{n} ( \int_{-\infty }^{0}+\int_{0}^{+\infty })
b_{j}( x) ^{\ast }i\frac{\partial }{\partial x}b_{j}( x) dx, \nonumber \\
J &=&i \sum_{j=1}^{n} \bar{b}_{j}( 0) ^{\ast }[ b_{j}( 0^{+}) -b_{j}( 0^{-})
] .
\end{eqnarray}

\subsubsection{Pure Scattering}

We take the singular potential to be the second quantization of the $\delta $%
-function written in an explicitly symmetric manner: 
\begin{eqnarray*}
\Upsilon =E_{ij}\,\bar{b}_{i}(0)^{\ast }\bar{b}_{j}(0).
\end{eqnarray*}
Then 
\begin{eqnarray*}
i\dot{\Psi} &=&(K_{0}+\Upsilon )\Psi \\
&=&\tilde{K}_{0}\Psi +\bar{b}_{i}(0)^{\ast }[i[b_{i}(0^{+})-b(0^{-})]+E_{ij}\bar{b}_{j}(0)]\Psi
\end{eqnarray*}
and again asking for the singular part (that is, the coefficient of $\bar{b}_{i}(0)^{\ast }$) to vanish leads to the boundary condition 
\begin{eqnarray*}
b_{i}(0^{-})\Psi = \left[\frac{1-\frac{i}{2}E}{1+\frac{i}{2}E} \right]_{ij}\,b_{j}(0^{+}) \Psi =S_{ij}\,b_{j}(0^{+})\Psi .
\end{eqnarray*}
We see that we have free propagation by translation along the incoming and outgoing wires. At the origin we have the boundary condition $b_{i}(0^{-})\Psi =S_{ij}\,b_{j}(0^{+})\Psi $.

\subsubsection{General Situation}

The fact that we are modeling a quantum field process traveling along the wire means that we may consider more general interactions than just scattering. In particular, we may include the action on the system due to the emission or absorption of the field quanta. To model this we now take 
\begin{eqnarray*}
\Upsilon =E_{ij}\,\bar{b}_{i}(0)^{\ast }\bar{b}_{j}(0)+E_{i0}\,\bar{b}_{i}(0)^{\ast }+E_{0j}\,\bar{b}_{j}(0)+E_{00}.
\end{eqnarray*}
Then 
\begin{eqnarray*}
i\dot{\Psi} &=&(K_{0}+\Upsilon )\Psi \\
&=&\{\tilde{K}_{0}+E_{i0}\bar{b}_{i}(0)+E_{00}\}\Psi \\
&+&\bar{b}_{i}(0)^{\ast
}\big\{ i[b_{i}(0^{+})-b_{i}(0^{-})]+E_{ij}\bar{b}_{j}(0)+E_{10} \big\} \Psi
\end{eqnarray*}
We once again ask that the final term involving $\bar{b}_{i}(0)^\ast$ (the singular part!) vanishes and this is equivalent to the algebraic condition: 
\begin{eqnarray*}
b_{i}(0^{-})\Psi =\left[ \frac{1-\frac{i}{2}E}{1+\frac{i}{2}E} \right]_{ij}\,b_{j}(0^{+})%
\Psi -i \left[ \frac{1}{1+\frac{i}{2}E} \right]_{ij}E_{j0}\Psi ,
\end{eqnarray*}
and, substituting in for the It\={o} coefficients $(\ref{L's})$, this is exactly the boundary conditions (\ref{eq:BC}), that is, $b_{i}(0^{-})\Psi =L_{i}\Psi +S_{ij}\,b_{j}(0^{+})\Psi $. Rearranging then leaves us with 
\begin{eqnarray*}
\dot{\Psi}=-i\tilde{K}_{0}\Psi -(\frac{1}{2}L_{i}^{\ast }L_{i}+iH)\Psi
-L_{i}^{\ast }S_{ij}b_{j}(0^{+})\Psi .
\end{eqnarray*}
From this we readily identify the desired form (\ref{eq:Greg_Ham}).

\section{Fermionic Models}

\label{sec:Fermi} In the Fermi analogue we encounter input processes $a_{i}(t) $ satisfying the singular canonical anti-commutation relations 
\begin{eqnarray}
\{ a_{i}( t) ,a_{j}( s) ^{\ast }\} =\delta _{ij}\delta ( t-s) ,  \label{SCAR}
\end{eqnarray}
with $\{ a_{i}( t) ,a_{j}( s) \} =0=\{ a_{i}( t) ^{\ast },a_{j}( s) ^{\ast }\} $. The appropriate Hilbert space to describe these objects is the Fermi Fock space $\mathfrak{F}_{-}$ consisting of vectors $\Psi $ with amplitudes $\langle \tau _{1},i_{1};\cdots ;\tau _{m},i_{m}|\Psi \rangle $ that are completely anti-symmetric under interchange of the labels $( \tau _{i},i_{j}) $. (By the exclusion principle, the amplitude vanishes if two labels are identical.) The Fermi annihilator is then defined by 
\begin{eqnarray*}
a_{i}( t) | \tau _{1},i_{1};\cdots ;\tau _{m},i_{m}\rangle =| i,t;\tau
_{1},i_{1};\cdots ;\tau _{m},i_{m}\rangle .
\end{eqnarray*}
On the domain of suitable test vector $\Psi \in \mathfrak{F}_{-}$, we then define the singular densities $a_{i}( 0^{\pm }) $ and $\bar{a} _{i}( 0) =\frac{1}{2}a_{i}( 0^{+}) +\frac{1}{2}a_{i}( 0^{-}) $.

The second quantization procedure is similar to the Bose case and we can immediately introduce the Fermi analogues of the time shift operators: 
\begin{eqnarray*}
K_{0} &=&\sum_{j=1}^{n}\int_{-\infty }^{\infty }dt\;a_{j}( t) ^{\ast }i\frac{%
\partial }{\partial t}a_{j}( t) , \\
\tilde{K}_{0} &=&\sum_{j=1}^{n}( \int_{-\infty }^{0}+\int_{0}^{\infty })
dt\;a_{j}( t) ^{\ast }i\frac{\partial }{\partial t}a_{j}( t) , \\
J &=&i\sum_{j=1}^{n}\bar{a}_{j}( 0) ^{\ast }[ a_{j}( 0^{+}) -a_{j}( 0^{-}) ]
.
\end{eqnarray*}

\subsection{Coupling to the System}

The theory of Fermionic quantum stochastic calculus was developed in the mid-1980s by Hudson and Applebaum \cite{Applebaum-Hudson}, \cite{Applebaum} for Fermi diffusions of even parity, and Hudson and Parthasarathy \cite{HP:UFBSC} for the general case. In applications to physical models we encounter restrictions on the type of coupling and dynamical evolutions, meaning that the full theory presented in the latter paper is too broad.

For instance, if the bath is an electron reservoir, then the specific issue that arises in practice is that the creation of an electron in the bath necessarily requires the removal of an electron from the system. This means
that the system must carry Fermi degrees of freedom. An example of a suitable Fermionic local Hamiltonian is 
\begin{eqnarray*}
\Upsilon (t)=\omega _{ij}a_{i}(t)^{\ast }a_{j}(t)+\eta _{\alpha \beta
}c_{\alpha }^{\ast }c_{\beta }\\
+\kappa _{\alpha j}c_{\alpha }^{\ast
}a_{j}(t)+\kappa _{\alpha j}^{\ast }a_{j}(t)^{\ast }c_{\alpha }
\end{eqnarray*}
where $c_{\alpha }$ are Fermionic modes of the system and the $\omega_{ij},\kappa _{\alpha j},\eta _{\alpha \beta }$ are constants. We require that the system modes satisfy anti-commutation relation $\{c_{\alpha },c_{\beta }^{\ast }\}=\delta _{\alpha \beta }$, $\{c_{\alpha },c_{\beta }\}=0=\{c_{\alpha }^{\ast },c_{\beta }^{\ast }\}$, and also anti-commute with the bath modes 
\begin{eqnarray*}
\{c_{\alpha },a_{i}(t)\}=\{c_{\alpha }^{\ast },a_{i}(t)\}=0 ,\\
\{c_{\alpha
},a_{i}(t)^{\ast }\}=\{c_{\alpha }^{\ast },a_{i}(t)^{\ast }\}=0.
\end{eqnarray*}

\subsubsection{Parity Restrictions}

We define the parity operator $\eta $ by 
\begin{eqnarray*}
\eta ( c_{\alpha }) &=&-c_{\alpha }, \\
\eta ( a_{i}( t) ) &=&-a_{i}( t) ,
\end{eqnarray*}
with $\eta ( XY) =\eta ( X) \eta ( Y) $ and $\eta ( X^{\ast }) =\eta ( X) ^{\ast }$. An operator $X $ on the joint system and bath space is said to be of even parity is $\eta ( X) =X$ and of odd parity if $\eta ( X) =-X$.

We note that the local Hamiltonian is of even parity, 
\begin{eqnarray*}
\eta ( \Upsilon ( t) ) =\Upsilon ( t) ,
\end{eqnarray*}
and this is a natural requirement for all physically realistic models. The most general type of local Hamiltonian that we shall consider will be of the form 
\begin{eqnarray*}
\Upsilon ( t) =E_{ij}a_{i}( t) ^{\ast }a_{j}( t) +a_{i}( t) ^{\ast
}E_{i0}+E_{0j}a_{j}( t) +E_{00}
\end{eqnarray*}
where the $E_{ij}$ are operators on the system space necessarily possessing the following definite parities

\begin{itemize}
\item[ ]  $E_{ij}$ - even

\item[ ]  $E_{i0}=( E_{0i}) ^{\ast }$ - odd

\item[ ]  $E_{00}$ - even.
\end{itemize}

We note that $a_{i}( t) ^{\ast }E_{i0}=-E_{i0}a_{i}( t) ^{\ast }$. As before, we wish to study the unitary 
\begin{eqnarray*}
U( t) =\vec{T}\exp \frac{1}{i}\int_{0}^{t}\Upsilon ( \tau ) d\tau
\end{eqnarray*}
which by construction should be of even parity and satisfy the cocycle relation with respect to the free translation on the Fermionic Fock space.

\subsubsection{Conversion to It\={o} Form}

As in the Bose case we encounter 
\begin{eqnarray*}
\lbrack a_{i}(t),U(t)]=-i\int_{0}^{t}[a_{i}(t),\Upsilon (s)]U(s)ds.
\end{eqnarray*}
The parities of the components of $\Upsilon $ are essential in computing $[a_{i}(t),\Upsilon (s)]$. For instance, using the anti-commutation relations and observing that $E_{jk}$ commutes with the bath modes $a_{i}(t)$, 
\begin{eqnarray*}
\lbrack a_{i}(t),E_{jk}a_{j}(s)^{\ast }a_{k}(s)]
&=&+E_{jk}a_{i}(t)a_{j}(s)^{\ast }a_{k}(s)\\ 
&&-E_{jk}a_{j}(s)^{\ast
}a_{k}(s)a_{i}(t) \\
&=&E_{jk}\{a_{i}(t),a_{j}(s)^{\ast }\}a_{k}(t) \\
&=&E_{jk}a_{k}(t)\delta (t-s).
\end{eqnarray*}
We in fact see that 
\begin{eqnarray*}
\lbrack a_{i}(t),\Upsilon (s)]=\delta _{ij}\{E_{jk}a_{k}(t)+E_{j0}\}\delta
(t-s)
\end{eqnarray*}
which is structurally identical to the Bose case. The Wick ordered form is therefore equivalent to (\ref{Wick QSDE}) 
\begin{eqnarray*}
\dot{U}(t)=a_{i}(t)^{\ast }(S_{ij}-\delta _{ij})U(t)a_{j}(t)+a_{i}(t)^{\ast
}L_{i}U(t) \\
-L_{i}^{\ast }S_{ij}U(t)a_{j}(t)-(\frac{1}{2}L_{i}^{\ast
}L_{i}-iH)U(t),
\end{eqnarray*}
The coupling operators $S,L,H$ take the same forms as in (\ref{S}, \ref{L's}) though we note that the carry the following definite parities listed in the table below

\begin{eqnarray*}
\begin{tabular}{l|l|l}
Parity & Even & Odd \\ \hline\hline
Bath Processes & $\Lambda _{ij}(t)$ & $A_{i}(t)$, $A_{i}(t)^{\ast }$ \\ 
System coefficients & $E_{ij}$, $E_{00}$ & $E_{i0}$, $E_{0j}$ \\ 
It\={o} coefficients & $S_{ij}$, $H$ & $L_{i}$%
\end{tabular}
\end{eqnarray*}

\subsection{It\={o} Form}

As in the Bose case we may introduce the integrated fields 
\begin{eqnarray*}
A_{i}( t) &=&\int_{0}^{t}a_{i}( s) ds,\;A_{i}( t) ^{\ast
}=\int_{0}^{t}a_{i}( s) ^{\ast }ds \\
\Lambda _{ij}( s) &=&\int_{0}^{t}a_{i}( s) ^{\ast }a_{j}( s) ds
\end{eqnarray*}
which are regular operators on the Fermi Fock space and which may be extended to operators on the joint system and bath space in the obvious manner. The operators $A_{i}( t) $ and $A_{i}( t) ^{\ast }$ are clearly odd,
while $\Lambda _{ij}( t) $ is even. They lead to a quantum It\={o} table that is exactly the same as the Bose case (\ref{Ito table}).

The It\={o} form of the QSDE is therefore 
\begin{eqnarray*}
dU(t) &=&\{(S_{ij}-\delta _{ij})d\Lambda _{ij}(t)+dA_{i}^{\ast
}(t)L_{i} \\
&& -L_{i}^{\ast }S_{ij}dA_{j}(t)-(\frac{1}{2}L_{i}^{\ast
}L_{i}+iH)dt\}U(t),
\end{eqnarray*}
and we note the change $dA_{i}^{\ast }L_{i}=-L_{i}dA_{i}^{\ast }$.

\subsubsection{Fermi Input-Output Relations}

An application of the It\={o} table shows that Fermi output fields defined by 
\begin{eqnarray*}
A_{i}^{\mathrm{out}}( t) =U( t) ^{\ast }A_{i}( t) U( t)
\end{eqnarray*}
will satisfy the differential relations 
\begin{eqnarray*}
dA_{i}^{\mathrm{out}}(t)=U( t) ^{\ast }S_{ij}U(t)dA_{j}( t) +U( t) ^{\ast
}L_{i}U( t) dt.
\end{eqnarray*}
While formally identical to the Bose case, we should emphasize that the Fermi input-output relation has the additional property that both sides of the relation are of odd parity.

\subsubsection{The Fermi Langevin Equations}

We again define $j_{t}(X)=U(t)^{\ast }XU(t)$. For $S=I$, the QSDE reduces to 
\begin{eqnarray*}
dU(t)=\{dA_{i}^{\ast }L_{i}-L_{i}^{\ast }dA_{i}-\{\frac{1}{2}L_{i}^{\ast
}L_{i}+iH)dt\}U(t)
\end{eqnarray*}
We now have 
\begin{eqnarray*}
dj_{t}(X) =& U(t)^{\ast }&XdU(t)+dU(t)^{\ast }XU(t)+dU(t)^{\ast }XdU(t) \\
=&U(t)^{\ast }&\{XdA_{i}^{\ast }L_{i}-XL_{i}^{\ast }dA_{i}-X(\frac{1}{2}%
L_{i}^{\ast }L_{i}+iH)dt \\
&+&L_{i}^{\ast }dA_{i}X-dA_{i}^{\ast }L_{i}X-(\frac{1}{2}L_{i}^{\ast
}L_{i}-iH)Xdt\\
&+& L_{i}^{\ast }dA_{i}XdA_{j}^{\ast }L_{j}\}U(t)
\end{eqnarray*}
and to proceed further we need to take into account the parity features of $X $.

For a given operator $Z$ we can write $Z$ as a sum of even and odd parts by setting $Z_{\mathrm{even}}=\frac{1}{2}Z+\frac{1}{2}\eta (Z)$ and $Z_{\mathrm{odd}}=\frac{1}{2}Z-\frac{1}{2}\eta (Z)$ to yield $Z=Z_{\mathrm{even}}+Z_{\mathrm{odd}}$ with $\eta (Z)=Z_{\mathrm{even}}-Z_{\mathrm{odd}}$. We see that 
\begin{eqnarray*}
ZdA_{i}^{\ast }(t)=dA_{i}^{\ast }(t)\eta (Z),\;dA_{i}(t)Z=\eta (Z)dA_{i}(t).
\end{eqnarray*}
The Langevin equation therefore becomes 
\begin{eqnarray*}
dj_{t}(X) &=&dA_{i}^{\ast }(t) \,j_{t}(\eta (X)L_{i}-L_{i}X)\\
&&+j_{t}(L_{i}^{\ast
}\eta (X)-XL_{i}^{\ast })\, dA_{i}(t) \\
&&+j_{t}(L_{i}^{\ast }\eta (X)L_{i}-\frac{1}{2}XL_{i}^{\ast }L_{i}-\frac{1}{2}L_{i}^{\ast }L_{i}X)\, dt\\
&& -i j_t ([X,H])dt
\end{eqnarray*}
and for $S\neq I$ this generalizes to 
\begin{eqnarray*}
dj_{t}(X) &=&j_{t}(S_{ki}^{\ast }XS_{kj}-\delta _{ij}X)\, d\Lambda _{ij}(t) \\
&&+dA_{i}^{\ast }(t) \, j_{t}(S_{ji}^{\ast }[\eta
(X)L_{j}\\
&& -L_{j}X])+j_{t}([L_{i}^{\ast }\eta (X)-XL_{i}^{\ast
}]S_{ij}) \, dA_{j}(t) \\
&&+j_{t}(L_{i}^{\ast }\eta (X)L_{i}-\frac{1}{2}XL_{i}^{\ast }L_{i}-\frac{1}{2}L_{i}^{\ast }L_{i}X) \ dt \\
&& -ij_t ([X,H]) \, dt.
\end{eqnarray*}

For even operators $X$ this is formally identical to the Bose Langevin
equation, however, for odd $X$ we have 
\begin{eqnarray*}
dj_{t}( X) &=&j_{t}( S_{ki}^{\ast }XS_{kj}-\delta _{ij}X) d\Lambda _{ij}( t)
\\
&&-dA_{i}^{\ast }( t) j_{t}(S_{ji}^{\ast }\{X,L_{j}\})-j_{t}(\{L_{i}^{\ast
},X\}S_{ij})dA_{j}( t) \\
&&-j_{t}(\frac{1}{2}\{X,L_{i}^{\ast }\}L_{i}+\frac{1}{2}L_{i}^{\ast
}\{L_{i},X\}+i[ X,H] )dt.
\end{eqnarray*}

\subsubsection{The Fermi Global Hamiltonian}

We can now state the global Hamiltonian $K$ for the Fermi case: 
\begin{eqnarray}
-i K\Psi =-i\tilde{K}_{0}\Psi -(\frac{1}{2}L_{i}^{\ast }L_{i}+iH)\Psi
-L_{i}^{\ast }S_{ij}a_{j}( 0^{+}) \Psi ,\nonumber \\
\end{eqnarray}
on the domain of suitable functions satisfying the boundary condition 
\begin{eqnarray}
a_{i}( 0^{-}) \Psi =L_{i}\Psi +S_{ij}\,a_{j}( 0^{+}) \Psi .
\end{eqnarray}

\section{Quantum Feedback Networks}
\label{sec:network}
\begin{figure}[htbp]
	\centering
		\includegraphics[width=0.40\textwidth]{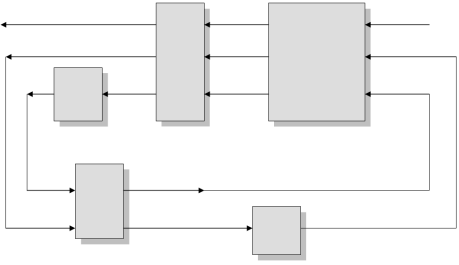}
	\caption{(color online) Several individual models (with inputs and outputs put into multiple blocks if necessary) are connected to form a quantum feedback network. the components
	form the set of vertices $\mathcal{V}$ of the network, and the input/output fields propagate along the edges. There will of necessity be external fields driving the network.}
	\label{fig:Ham_network}
\end{figure}

A general quantum feedback network consists of a direct graph with vertices $\mathcal{V}$ and edges $\mathcal{E}$, see Fig. \ref{fig:Ham_network}. At each vertex we have a quantum mechanical system described by the triple $\left( S_{v},L_{v},H_{v}\right) $. 

To describe the open-loop model, we may form the concatenation $\left( S ,L,H\right) $ where 
\begin{eqnarray*}
S=\left[ 
\begin{array}{ccc}
S_{1} & 0 & \cdots \\ 
0 & S_{2} & \cdots \\ 
\vdots & \vdots & \ddots
\end{array}
\right] ,\,L=\left[ 
\begin{array}{c}
L_{1} \\ 
L_{2} \\ 
0
\end{array}
\right] ,\,H=\sum_{v\in \mathcal{V}}H_{v}.
\end{eqnarray*}
The closed loop arrangement comes from feeding output fields in as input fields as indicated in the network graph. The global Hamiltonian $K$ for the network will take the form \cite{GoughJamesIEEE09} 
\begin{eqnarray*}
-iK\Psi =-i\sum_{e\in \mathcal{E}}\tilde{K}_{e}\Psi -\sum_{v\in \mathcal{V}}\left( \frac{1}{2}L_{v}^{\ast }L_{v}+iH_{v}\right) \Psi \\
-\sum_{v\in 
\mathcal{V}}L_{v}^{\ast }S_{v}b_{v}\left( +\right) \Psi
\end{eqnarray*}
where $b_{v}\left( +\right) $ is the vector of incoming annihilator densities evaluated immediately before vertex $v\in \mathcal{V}$. This must be supplemented by the set of boundary conditions 
\begin{eqnarray*}
b_{v}\left( -\right) \Psi =S_{v}b_{v}\left( +\right) \Psi +L_{v}\Psi
\end{eqnarray*}
for each vertex $v\in \mathcal{V}$. Here $\tilde{K}_{e}$ is the generator of free translation along each particular edge $e\in \mathcal{E}$. A detailed account may be found in \cite{GoughJamesCMP09}.

\subsection{The Series Product}

We illustrate the method next with a derivation of the series product of \cite{GoughJamesIEEE09} using a general argument introduced in \cite{GoughJamesCMP09}. The basic set up is sketched in Figure 1.

\begin{figure}[htbp]
	\centering
		\includegraphics[width=0.40\textwidth]{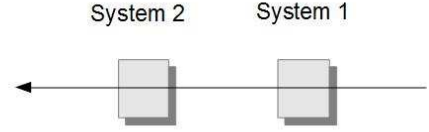}
	\caption{(color online) Two systems $(S^{(1)}, L^{(1)},H^{(1)})$ and $(S^{(2)}, L^{(2)},H^{(2)})$ connected in series.}
	\label{fig:Ham_series}
\end{figure}

We may model a pair of systems in cascade by specifying the local Hamiltonians through the triples $\left( S^{\left( i\right) },L^{\left( i\right) },H^{\left( i\right) }\right) $ for $i=1,2$. The positions of the systems are $t_{1}$ and $t_{2}$ respectively. The global Hamiltonian is then 
\begin{eqnarray*}
-iK\Psi &=&-i\left( \int_{-\infty
}^{t_{2}}+\int_{t_{2}}^{t_{1}}+\int_{t_{1}}^{\infty }\right) b_{j}\left(
t\right) ^{\ast }i\frac{\partial }{\partial t}b_{j}\left( t\right) \\
&&-(\frac{1}{2}L_{j}^{\left( 1\right) \ast }L_{j}^{\left( 1\right)
}+iH^{\left( 1\right) })\Psi -L_{j}^{\left( 1\right) \ast
}S_{jk}^{(1)}b_{k}\left( t_{1}^{+}\right) \Psi \\
&&-(\frac{1}{2}L_{j}^{\left( 2\right) \ast }L_{j}^{\left( 2\right)
}+iH^{\left( 2\right) })\Psi -L_{j}^{\left( 2\right) \ast
}S_{jk}^{(2)}b_{k}\left( t_{2}^{+}\right) \Psi
\end{eqnarray*}
with boundary conditions 
\begin{eqnarray*}
b_{j}\left( t_{1}^{-}\right) \Psi &=&S_{jk}^{\left( 1\right) }b_{k}\left(
t_{1}^{+}\right) \Psi +L_{j}^{\left( 1\right) }\Psi , \\
b_{j}\left( t_{2}^{-}\right) \Psi &=&S_{jk}^{\left( 2\right) }b_{k}\left(
t_{2}^{+}\right) \Psi +L_{j}^{\left( 2\right) }\Psi .
\end{eqnarray*}

To obtain the instantaneous feedforward limit we take $t_{1}-t_{2}\rightarrow 0$. Denoting the common limit as $t_{0}$ then 
\begin{eqnarray*}
b_{j}\left( t_{0}^{-}\right) \Psi &=&S_{jk}^{\left( 2\right)
}\lim_{t_{2}\rightarrow t_{0}}b_{k}\left( t_{2}^{+}\right) \Psi
+L_{j}^{\left( 2\right) }\Psi \\
&=&S_{jk}^{\left( 2\right) }\lim_{t_{1}\rightarrow t_{0}}b_{k}\left(
t_{1}^{-}\right) \Psi +L_{j}^{\left( 2\right) }\Psi \\
&=&S_{jk}^{\left( 2\right) }\lim_{t_{1}\rightarrow t_{0}}\left\{
S_{jk}^{\left( 1\right) }b_{k}\left( t_{1}^{+}\right) \Psi +L_{j}^{\left(
1\right) }\Psi \right\} +L_{j}^{\left( 2\right) }\Psi \\
&=&S_{jk}^{\left( 2\right) }\left\{ S_{jk}^{\left( 1\right) }b_{k}\left(
t_{0}^{+}\right) \Psi +L_{j}^{\left( 1\right) }\Psi \right\} +L_{j}^{\left(
2\right) }\Psi .
\end{eqnarray*}
The global Hamiltonian then reduces to 
\begin{eqnarray*}
-iK\Psi &=&-i\left( \int_{-\infty }^{t_{0}}+\int_{t_{0}}^{\infty }\right)
b_{j}\left( t\right) ^{\ast }i\frac{\partial }{\partial t}b_{j}\left(
t\right) \Psi \\
&&-(\frac{1}{2}L_{j}^{\left( 1\right) \ast }L_{j}^{\left( 1\right)
}+iH^{\left( 1\right) })\Psi -L_{j}^{\left( 1\right) \ast
}S_{jk}^{(1)}b_{k}\left( t_{0}^{+}\right) \Psi \\
&&-(\frac{1}{2}L_{j}^{\left( 2\right) \ast }L_{j}^{\left( 2\right)
}+iH^{\left( 2\right) })\Psi -L_{j}^{\left( 2\right) \ast
}S_{jk}^{(2)}b_{k}\left( t_{0}^{+}\right) \Psi .
\end{eqnarray*}
Substituting in gives 
\begin{eqnarray*}
-iK\Psi &=& -i\left( \int_{-\infty }^{t_{0}}+\int_{t_{0}}^{\infty }\right)
b_{j}\left( t\right) ^{\ast }i\frac{\partial }{\partial t}b_{j}\left(
t\right) \Psi \\
&&-(\frac{1}{2}L_{j}^{\ast }L_{j}+iH)\Psi -L_{j}^{\ast
}S_{jk}b_{k}\left( t_{0}^{+}\right) \psi
\end{eqnarray*}
with boundary condition 
\begin{eqnarray*}
b_{j}\left( t_{0}^{-}\right) \Psi =S_{jk}b_{k}\left( t_{0}^{+}\right) \Psi
+L_{j}\Psi
\end{eqnarray*}
where we have 
\begin{eqnarray*}
S &=&S^{\left( 2\right) }S^{\left( 1\right) },L=L^{\left( 2\right)
}+S^{\left( 2\right) }L^{\left( 1\right) }, \\
H &=&H^{\left( 1\right) }+H^{\left( 2\right) }+\mathrm{Im}\left\{ L^{\left(
2\right) \ast }S^{\left( 2\right) }L^{\left( 1\right) }\right\} .
\end{eqnarray*}

The rule 
\[\left( S^{\left( 2\right) },L^{\left( 2\right) },H^{\left(
2\right) }\right) \vartriangleleft \left( S^{\left( 1\right) },L^{\left(
1\right) },H^{\left( 1\right) }\right) =\left( S,L,H\right) 
\]
determined
above is referred to as the \emph{series product} of the cascaded system 
\cite{GoughJamesIEEE09}.

\subsection{General Feedback Reduction}

More generally, the reduced model obtained from a concatenation $(S,L,H)$ obtained by eliminating the edge $(r_{0},s_{0})$ is shown in Figure 2.

\begin{figure}[htbp]
	\centering
		\includegraphics[width=0.40\textwidth]{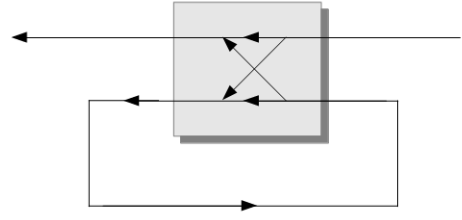}
	\caption{(color online) The internal line is fed back in as input making an algebraic loop.}
	\label{fig:Ham_feedback}
\end{figure}

By a similar
argument, it is readily seen to be determined by the operators $(S^{\mathrm{red}},L^{\mathrm{red}},H^{\mathrm{red}})$ where \cite{GoughJamesCMP09} 
\begin{eqnarray*}
S_{sr}^{\mathrm{red}} &=&S_{sr}+S_{sr_{0}}\left( 1-S_{s_{0}r_{0}}\right)
^{-1}S_{s_{0}r}, \\
L_{s}^{\mathrm{red}} &=&L_{s}+S_{sr_{0}}\left( 1-S_{s_{0}r_{0}}\right)
^{-1}L_{s_{0}}, \\
H^{\mathrm{red}} &=&H+\sum_{s\mathrm{:\,output\,edge}}\mathrm{Im}L_{s}^{\ast
}S_{sr_{0}}\left( 1-S_{s_{0}r_{0}}\right) ^{-1}L_{s_{0}},
\end{eqnarray*}
We comment that the same rule applies to the Fermi case as well, and in particular that the series product, and feedback reduction rule ( preserve the correct parity in table above.

\section{Conclusion}

We have shown that the standard Markov models for quantum mechanical systems driven by quantum inputs may be formulated as the free translation of an indefinite number of indistinguishable quanta along a one-dimensional wire
with a singular localized interaction at the point which the system is placed. The approach works equally well for Boson and Fermion quanta. More generally we may consider quantum network models with quanta propagating along the edges and localized quantum mechanical systems located at the vertices. 

Under physically motivated assumptions on the parity of the coupling operator coefficients, the Fermi analogue leads to identical equations for the series product for cascaded and direct feedback situations, and more generally for the feedback reduction formula for closed-loop networks involving general feedback relations.

\section*{Acknowledgments}

It is a great pleasure to thank Matthew James and Hendra Nurdin for several
stimulating discussions on quantum feedback networks. He is also
indebted to Alexei Iantchenko for pointing out the work of Albeverio and
Kurasov for the first-quantization scattering problem. The support of the UK
Engineering and Physical Sciences research council under grants EP/H016708/1
and EP/G020272/1 is gratefully acknowledged.

\end{document}